\documentclass[american,aps,pra,reprint, superscriptaddress]{revtex4-1}
\usepackage{color}
\usepackage{times}
\usepackage{subfig}
\usepackage{bm}
\usepackage{graphicx}
\usepackage{graphics}
\DeclareGraphicsExtensions{.pdf,.png,.jpg}
\usepackage{amsbsy}
\usepackage{amsmath}
\usepackage{amsfonts}
\usepackage{amsthm}
\usepackage{float}
\usepackage{amssymb}
\usepackage{textcomp}
\usepackage{mathtools}
\usepackage[subnum]{cases}
\usepackage{enumitem}

\DeclareMathOperator{\Tr}{Tr}
\definecolor{purple(html/css)}{rgb}{0.5, 0.0, 0.5}
\newcommand{\ket}[1]{| #1 \rangle}
\newcommand{\bra}[1]{\langle #1 |}
\newcommand{\braket}[2]{\langle #1 | #2 \rangle}
\newcommand{\ketbra}[2]{| #1 \rangle \langle #2 |}

\begin{document}

\title{Quantum Coherence, Coherent Information and Information Gain in Quantum Measurement}

\author{Gautam Sharma}
\email{gautam.oct@gmail.com}
\affiliation{Quantum Information and Computation Group, Harish-Chandra Research Institute,\\ Homi Bhabha National Institute, Allahabad, 211019, India}

\author{Sk Sazim}
\email{sk.sazimsq49@gmail.com}
\affiliation{Quantum Information and Computation Group, Harish-Chandra Research Institute,\\ Homi Bhabha National Institute, Allahabad, 211019, India}

\author{Arun K. Pati}
\email{akpati@hri.res.in}
\affiliation{Quantum Information and Computation Group, Harish-Chandra Research Institute,\\ Homi Bhabha National Institute, Allahabad, 211019, India}

\begin{abstract}
 A measurement is deemed successful, if one can maximize the information gain by the measurement apparatus. Here, we ask if quantum coherence of the system imposes a limitation on the information gain during quantum measurement. First, we argue that the information gain in a quantum measurement is nothing but the coherent information or the distinct quantum information that one can send from the system to apparatus. We prove that the maximum information gain from a pure state, using a mixed apparatus is upper bounded by the initial coherence of the system. Further, we illustrate the measurement scenario in the presence of environment. 
 We argue that the information gain is upper bounded by the entropy exchange between the system and the apparatus. Also, to maximize the information gain, both the initial coherence of the apparatus, and the final entanglement between the system and apparatus should be maximum. Moreover, we find that for a fixed amount of coherence in the final apparatus state the more robust apparatus is, the more will be the information gain.
\end{abstract}

\maketitle
\section{Introduction}
Unlike the Schr\"{o}dinger evolution, quantum measurement process is a non-unitary evolution of a physical system which gives information about the physical quantity of interest. Every measurement process has one key 
element: it establishes correlations between the system and the apparatus \cite{RevModPhys.29.454}. These correlations are both classical and quantum in nature. The quantum part of the correlations is presumed to be entanglement \cite{VonNeumann,PhysRevLett.88.017901,ccorr,discordreview}. The more entanglement between the apparatus and system is, the more successful will be the measurement \cite{measurementWheeler,Busch2003,2011NaPho...5..222G,RevModPhys.86.307,PhysRevA.97.062308}. 

Recently, it has been shown that to create quantum correlations between two subsystems, quantum coherence may 
provide the required initial resource \cite{PhysRevLett.115.020403, PhysRevLett.116.160407}. In such processes, the amount of generated entanglement will always be less than the initial coherence in one of the subsystems. In a generic measurement process, this finding may help one to extract information present in the form of coherence of the system under consideration. Moreover, it has been shown that quantum coherence is connected with other important resource theories as well \cite{1751-8121-51-41-414006,1367-2630-20-5-053058}. Furthermore, there exists another connection between the quantum coherence and entanglement. For a bipartite system, the coherence of any subsystem and the entanglement of the total system respect a complementarity relation \cite{RIS_0}. We wonder whether this finding may assist in reducing the effect of environment during a measurement process. These results alludes us to investigate the role of coherence in quantum measurement setting. 

If the system is prepared in one of the eigenstate of the observable, then measurement will yield the eigenvalue of the observable with certainty. However, if the system is in a superposition of eigenstates of the observable, then the measurement outcome becomes probabilistic. Therefore, it is the superposition or coherence content of the initial system that is responsible for randomness in the measurement process. Then, a natural question comes to mind is that if coherence also decides the information gain in a quantum measurement process.

Though, information gain as well as information loss in a measurement process, have been studied in the past \cite{ccorr,RevModPhys.74.197,PhysRevLett.90.050401,PhysRevLett.120.160501, Xi2013}, they are far from being complete. One may think that the more entangled the apparatus and system are, the more will be the information gain. However, in Ref.\cite{PhysRevLett.90.050401} it was shown that the maximal possible information, we can extract, is exactly equal to the classical correlations developed during the measurement process , which is always upper bounded by the entanglement. 
As quantum coherence is an important resource and closely connected with entanglement, we would like to investigate whether we can harness it to improve the information gain (or, minimize information loss) in a measurement process. This notion is physically appealing since one has better control over the coherence of the system than the final entanglement.

In this paper, we delve on how the quantum coherence of the system and the apparatus affect the information gain during a quantum measurement.
In this work, we study the effect of quantum coherence on the information gain in a measurement. We consider a particular measurement setting and find that the initial coherence of the system upper bounds the information gain. We also show that the information gain is equal to the coherent information from the final state of system to the apparatus. Further, we show that the information gain is always less than the entropy exchange between the apparatus and the system during the measurement.  We prove a complementarity relation between the disturbance caused to the apparatus, information gain and the coherence of the final state of the apparatus. We also find that in the presence of environment, the more is the initial coherence of the apparatus, the more will be the information gain. Thus, the quantum coherence of the system and apparatus play a decisive role in information gain during a quantum measurement.

The paper is organized as follows: We begin by introducing the relevant concepts required to describe our main results in a prelude. In Sec. \ref{sec2}, we discuss the role played by the coherence of the system and coherent information in determining the information gain. We extend our analysis to generic measurement scenarios by including the effect of environment in Sec.\ref{sec3}. We also discuss that the entropy exchange between the system and apparatus, and disturbance caused to the apparatus plays a pivotal role in determining the information gain. Finally, we conclude in Sec. \ref{sec4}.

{\em Prelude}.--
Before going to the main results, here, we introduce the concept of the quantum coherence and classical correlations. Quantum coherence arises due to the superposition principle. 
Unlike other resources, quantum coherence is a basis dependent quantity, 
and we should fix a particular basis, $\{\ket{i}\}$ ($i=1\ldots d$) in $\mathcal{H}_d$. The diagonal density matrices in this basis are called incoherent states, denoted by the set $\mathcal{I}$  \cite{cohereR1,cohereC4,cohereC9, cohereD7}. An incoherent operation takes an incoherent state to another incoherent state.

Any proper measure of the coherence $C$ must satisfy the following conditions \cite{cohereC4,cohereR}: (i) It should not increase under any incoherent operation; (ii) It is non-increasing under selective incoherent operations on average, and (iii) It does not increase under mixing of quantum states.
There exist a numerous coherence measures in the literature but for our purpose here, we focus on the relative entropy of coherence. It is defined as $C_{R}(\rho):=S(\rho||\rho^{D})$ \cite{cohereC4,cohereR}, where $S(\rho||\sigma)=-{\rm Tr}[\rho \log \sigma] -S(\rho)$ is the quantum relative entropy and $\rho^{D}=\sum_i\bra{i}\rho\ket{i}\ketbra{i}{i}$ is a completely dephased version of the state $\rho$ in the basis $\{\ket{i}\}$.

The Holevo quantity is the maximum possible information one can extract from a quantum ensemble $\{p_i,\rho_i\}$ \cite{chuang}  and it is 
defined as 
\begin{equation}
 \chi (\{p_i,\rho_i\})= \sum_ip_i S(\rho_i||\rho),
\end{equation}
where $\rho=\sum_ip_i\rho_i$. 
The classical correlations \cite{ccorr} and the Holevo quantity are related concepts \cite{zwolak}. A bipartite state will have classical correlations if after application 
of rank one Projective Operator Valued Measurement (POVM) (i.e., entanglement breaking operation) on one of the parties will transform the state to either classical-quantum (CQ) or quantum classical (QC) states \cite{zwolak}, i.e.,  $\rho_{CQ}=\sum_ip_i\ketbra{i}{i}_C\otimes\rho_Q^i$ or  $\rho_{QC}=\sum_ip_i\rho_Q^i\otimes\ketbra{i}{i}_C$, where $\rho^i$ are not orthogonal.  
The classical correlations \cite{ccorr} of a quantum state $\rho_{SA}$ is given by 
\begin{equation}
 J=\sup_{\pi_j}\left[S(\rho_{S})-\sum_jp_jS(\rho_{S|\pi_j^A})\right],
\end{equation}
where $\rho_{S|\pi_j^A}$ are the post measurement states with probability $p_i$ due to the application of 
projective measurements ($\{\pi_j\}$) on the part $A$ of $\rho_{SA}$. 
Then, the average post measurement state will be of the form of QC state. 
Therefore, the classical correlations of the state $\rho_{SA}$ is equivalent to the maximum possible 
mutual information of $\rho_{QC}$, i.e., $J=I(\rho_{QC})=S(\sum_ip_i\rho_i)-\sum_ip_iS(\rho_i)$, hence, the classical 
correlations is nothing but the Holevo quantity \cite{zwolak}.

%

\section{Quantum coherence of system and Information Gain}\label{sec2}
Here, we ask whether  the quantum coherence of the system plays a role in determining the information gain during a measurement process.

To answer the above query, we study the quantum measurement process in which the apparatus is initially in a mixed state.  We adopt and analyze the  measurement process described in Ref.\cite{PhysRevLett.90.050401}, where the initial system is in a  pure state $\ket{\Psi_S}=\sum_is_i\ket{s_i}$ and the initial state of the apparatus is a mixed state as given by  $\rho_A=\sum_ia_i\ketbra{a_i}{a_i}$, with $a_i$ as eigenvalues of $\rho_A$ and $\{\ket{a_i}\}$ as the eigenbasis. The measurement process can be described by a unitary operator, $U$, which acts on the system and apparatus together, such that
\begin{align*}
U(\ketbra{s_i}{s_j} \otimes \rho_A )U^{\dagger} = \ketbra{s_i}{s_j} \otimes \rho_{ij}.
\end{align*}
\noindent Here, we have assumed that the state $\ket{s_j}\ket{a_i}$ of system and apparatus is transformed into, $\ket{s_j}\ket{\tilde{a}_{ij}}$. In an ideal measurement process, the final states of apparatus corresponding to different system states are orthogonal to each other and can be distinguished perfectly. But more often, this is not the case and the information extracted is not maximum. As we are trying to extract maximum information of the system from the apparatus, we assume the apparatus states $\ket{\tilde{a}_{ij}}$ corresponding to different system states to be orthogonal to each other, i.e. , $\braket{\tilde{a}_{ij}}{\tilde{a}_{ik}}= \delta_{jk}$ . This unitary evolution allows us to develop correlation between the system and the apparatus which is given by 
\begin{align*}
\ket{\Psi}_S\bra{\Psi}\otimes\rho_{A}\rightarrow U(\ket{\Psi}_S\bra{\Psi}\otimes\rho_{A})U^{\dagger}=\rho_{S'A'}.
\end{align*}

The final joint state in general is an entangled state as given by \cite{PhysRevLett.90.050401}
\begin{align*}
\rho_{S'A'}= \sum_i |s_i|^2\ket{s_i}\bra{s_i} \otimes \rho_{ii} + \sum_{i\neq j}s_is^*_j\ket{s_i}\bra{s_j} \otimes \rho_{ij}.
\end{align*}
Note that the first term on the right carries the extractable information due to measurement. To extract the  maximum possible information from the system, we should be able to distinguish between the different post measurement apparatus states $\rho_{ii}$ precisely. The maximum amount of accessible information from the apparatus is given by the Holevo quantity,
\begin{align}
I_m = S(\sum_i|s_i|^2\rho_{ii})-\sum_i|s_i|^2S(\rho_{ii}).
\label{hol-extr}
\end{align}
This quantity can also be identified with the classical correlations of the state $\rho_{S'A'}^c=\sum_i |s_i|^2\ket{s_i}\bra{s_i} \otimes \rho_{ii}$. Note also that the state $\rho_{S'A'}^{c}$ is only classically correlated although the final state $\rho_{S'A'}$ has non-zero entanglement. It was argued that the maximum possible information gain during the process may not be identified by the entanglement developed in the system and apparatus. This can be understood by the following inequality \cite{PhysRevLett.90.050401}
\begin{align}
E_R(\rho_{S'A'})\geq I_m. 
\label{ent_ved}
\end{align}
where  $E_R(\rho_{AB})=\min_{\sigma_{AB}}S(\rho_{AB}||\sigma_{AB})$, is the relative entropy of entanglement \cite{PhysRevLett.78.2275,PhysRevA.57.1619} and $\sigma_{AB}$ is a separable state.
 
Now, we will show that during a quantum measurement the information gain is actually equal to the coherent information for the system and the apparatus state $\rho_{A'S'}$. Using Eq.(\ref{hol-extr}), we can write the information gain as
\begin{align}
I_m &= S(\sum_i|s_i|^2\rho_{ii})-\sum_i|s_i|^2S(\rho_{ii}) \nonumber \\
&=S(\rho_{A'})- S(\rho_A) \nonumber\\ 
& =S(\rho_{A'})- S(\rho_{S'A'}) \nonumber \\
 & = I_c(S'\rangle A'),
 \label{side_Im}
\end{align}
where we have used the fact that $S(\rho_{S'A'})=S(\rho_A)=S(\rho_{ii})$, as $\rho_{SA}$ evolves unitarity and $\rho_{S}$ is pure. The quantity, $I_c(S'\rangle A')=S(\rho_{A'})-S(\rho_{S'A'})$ is the coherent information of the final state from the system to the final state of the apparatus \cite{PhysRevA.54.2629, wilde_2013}. Hence, the extractable information is exactly equal to the amount of distinct quantum information, one may send from the system to the apparatus via measurement. Therefore, this finding suggest that the extractable information is actually of quantum origin even though it is captured by the classical correlations. 
Next, we will prove a trade-off relation for the information gain, the coherence of the apparatus states and the mixedness of the initial apparatus state. If we fix a basis for the apparatus, then $C_R(\rho_{A'})=S(\rho_{A'}^D)-S(\rho_{A'})$. Using Eq.(\ref{side_Im}), we find that 
\begin{align}
 I_m+C_R(\rho_{A'})+S(\rho_A)\leq \log N,
\end{align}
where $S(\rho_A)$ denotes mixedness of the apparatus and $N$ is its dimension. This relation tells us that 
to maximize information gain the coherence in the final apparatus as well as the mixedness of the 
initial state of the apparatus should be as minimum as possible. This relation is stronger than the one given in Ref.\cite{PhysRevLett.90.050401}, which reads is 
\begin{align}\label{IG-mixedness}
I_m+S(\rho_A)\leq \log N,
 \end{align}
i.e., the more mixedness in the apparatus state, the more difficult will be to extract the information.

Now, we ask how does the initial coherence of the system govern the information gain? We will actually prove that the initial coherence of the system puts an upper bound on the information gain during the measurement. 
We can write the initial coherence of the system state in the basis $\{\ket{s_i}\}$ as,
\begin{align*}
C_R(\ket{\Psi}\bra{\Psi})= -\sum_i|s_i|^2\log |s_i|^2,
\end{align*}
which is exactly same as the maximum information possible in the measurement.
This quantity is always greater than the information gain, i.e., $C_R(\ket{\Psi}\bra{\Psi}) \geq I_m$. 
 The above inequality comes from the fact that $S(\sum_i|s_i|^2\rho_{ii})\leq -\sum_i|s_i|^2\log |s_i|^2+\sum_i|s_i|^2S(\rho_{ii})$ \cite{chuang}. Hence, we get 
 \begin{align}\label{cohlimitsinfo}
 C_R(\ket{\Psi}\bra{\Psi}) \geq I_m = I_c(S'\rangle A').
 \end{align}
Therefore, the extractable information is upper bounded by the coherence of the initial state of the system. 
The reason why the measurement process cannot extract maximum information, ($C_R(\ket{\Psi}\bra{\Psi})$), is that the apparatus is mixed. This can be understood by the complementarity relation between the information gain and mixedness of the system given in Eq.(\ref{IG-mixedness}). 
%

As the coherence of the initial systems can be better control than the entanglement in the final state $\rho_{A'S'}$, our relation given in Eq.(\ref{cohlimitsinfo}) may be more useful operationally than Eq.(\ref{ent_ved}). In the limiting case of pure apparatus, the maximum information gain is equal to the initial coherence of the system and also to the entanglement developed between the system and the apparatus \cite{PhysRevLett.90.050401}.

\section{Coherence, Entanglement and Information Gain in presence of environment}\label{sec3}
We can generalize the above measurement scenario to a more realistic one by considering the environment also. Note that the environment here will not cause decoherence to the system rather we will use it to purify the mixed state of the apparatus, and hence, the initial joint state of the apparatus and environment can be expressed as $\ket{\Psi_{AE}}=\sum_i\sqrt{a_i}\ket{a_i}\ket{e_i}$. The transformation of the complete state is now given by 
\begin{align*}
\sum_is_i\ket{s_i}\otimes \ket{\Psi_{AE}} \longrightarrow\sum_is_i\ket{s_i} \ket{\Psi^i_{AE}},
\end{align*}
where $\ket{\Psi^i_{AE}}=\sum_j\sqrt{a_j}\ket{\tilde{a}_{ij}}\ket{e_j}$. One can easily see that tracing out the environment gives the same post-measurement state of the apparatus, which was obtained using only the mixed apparatus. Now, the final joint state of system and apparatus is obtained by the tracing out the environment, i.e., $\rho_{S^{\prime}A^{\prime}}=\sum_{ij}s_is^{*}_j(\sum_k\bra{e_k}\ket{\Psi^i_{AE}}\bra{\Psi^j_{AE}}\ket{e_k})\otimes\ket{s_i}\bra{s_j}$. On tracing out the environment we should get the the same post-measurement joint state of system and apparatus. Therefore, we should have $\sum_k\bra{e_k}\ket{\Psi^i_{AE}}\bra{\Psi^j_{AE}}\ket{e_k}=\rho_{ij}$. Note that during the whole process, the state of the environment remains unaffected. For a better understanding of the Eq.(\ref{IG-mixedness}), we can look at the initial joint state $\ket{\Psi_{AE}}$ as a bigger pure apparatus which can be used to extract the maximum information from the system. However, as we have no access to the environment, we loose out some information. 

In this section, we will prove that the information gain by the apparatus can never exceed the entropy exchange between the apparatus and the system during the measurement process. Also, we can prove a new complementarity relation for the information gain and the coherence content of the final state of the system after measurement. 

Note that when the system and the apparatus interact unitarily, the apparatus undergoes a noisy quantum evolution ($\Phi$) as given by 
\begin{align*}
&\rho_{A} \longrightarrow \Phi(\rho_A)
=\sum_{\mu}A_{\mu}\rho A_{\mu}^{\dagger}\\
&= \Tr_{SE}[U(\ketbra{\Psi_S}{\Psi_S}\otimes \ketbra{\Psi_{AE}}{\Psi_{AE}})U^{\dagger}] 
=\sum_i|a_i|^2 \rho_{ii}.
\end{align*}
Now, the coherent information for the state $\rho_{A}$ and channel $\Phi$ is $I_c(A\rangle E)=S(\Phi(\rho_{A}))-S(\Phi \otimes \mathcal{I}\ket{\Psi}_{AE}\bra{\Psi})=S(\rho_{A'})-S(\rho_{A'E})$.
This quantity represents how much entanglement is retained by the apparatus and the environment after the apparatus interacts with the system. Since $\ket{\Psi_{SAE}}$ evolves unitarily, we have $S(\rho_{A'E})=S(\rho_{S'})$. Using the inequality $I_c(A\rangle E)\leq S(\rho_A)$, we have $S(\rho_{A'})-S(\rho_{A'E})\leq S(\rho_{A})$ and hence
\begin{align*}
I_m=S(\rho_{A'})-S(\rho_{A})\leq S(\rho_{A'E}) .
\end{align*}
Since $S(\rho_{A'E})=S(\rho_{S'})=S_e$ is the entropy exchange between the apparatus and the system \cite{PhysRevA.54.2614, RIS_034t}, we have 
\begin{align}
I_m\leq S_e.
\end{align}
The entropy exchange $S_e$ is intrinsic property of the apparatus  and the dynamical map $\Phi$ that the apparatus undergoes. Also, this represents the entropy increase of the system state if it is initially in the pure state. Therefore, one can say that the information gain by the apparatus can never exceed the entropy exchange between the apparatus and the system during the measurement. 
Next, one may ask is there any trade-off relation between the information gain and the coherence of the final state of the system in case of non-ideal measurement. Using the above relation, we find that indeed they satisfy a complementarity relation which is given by  
\begin{align}
I_m +C_R(\rho_{S'})\leq \log M,
\end{align}
where $M$ is the dimension of the system. One may wonder whether the evolved state of the system will have non-zero coherence in the basis $\{\ket{s_i}\}$. However, we note that after the interaction, we have $\rho_{S'A'}=\sum_{ij}s_is_j^*\ket{s_i}\bra{s_j}\otimes \rho_{ij}$, which yields 
\begin{align*}
 \rho_{S'}&=\sum_{ij}s_is_j^*{\rm Tr}[\rho_{ij}]\ket{s_i}\bra{s_j}\\
 &=\sum_i |s_i|^2\ket{s_i}\bra{s_i}+\sum_{i\neq j}s_is_j^*{\rm Tr}[\rho_{ij}]\ket{s_i}\bra{s_j},
\end{align*}
as $Tr(\rho_{ii})=1$. Clearly, $\rho_{S'}$ is not diagonal in $\{\ket{s_i}\}$ basis unless $\Tr(\rho_{ij})=\delta_{ij}$ and therefore, has a non-zero coherence, i.e., $C_R(\rho_{S'})\neq 0$.

Now, we will show the trade off relations between the entanglement, coherence, and information gain by the apparatus in the presence of environment. It was shown by Vedral \cite{PhysRevLett.90.050401}, that the entanglement between apparatus and environment and the information from the measurement obey the following the complementarity relation 
\begin{align}
E_R(\rho_{A^{\prime}E}) + I_m \leq S(\rho_{A^{\prime}}), 
\label{env_ent}
\end{align}
\noindent where $\rho_{A^{\prime}}$ is the final state of the apparatus. It was argued that for extracting larger information from the system, the final entanglement between the apparatus and the environment should be less. From Eq.(\ref{env_ent}), one can obtain the following complementarity relation involving the final entanglement between apparatus and environment, extractable information, and final coherence of the apparatus.
\begin{align}\label{compL1}
E_R(\rho_{A^{\prime}E}) + I_m + C_R(\rho_{A^{\prime}}) \leq \log N.
\end{align}
 This equation tells that for extracting larger information, we want both the final coherence of the apparatus and the final entanglement between apparatus and environment to be small. However, prima facie, this is not clear whether at the same time the final coherence of the apparatus and the final entanglement between apparatus and environment will be small. We will use a recently introduced complementarity relation between coherence and entanglement to provide better intuition in this regard. 
 
 For the bipartite system $\ket{\Psi_{AE}}$ we have $C_R(\rho_A)+E(\rho_{AE})\leq \log N$ \cite{RIS_0}, where $E(\cdot)$ may be any bona-fide measure of entanglement. Although this relation is basis independent, but it does not properly reveal much information about the dual nature of the two resources. However, without loss of generality, if one can choose a preferred basis in which the coherence of any sub-system is maximum (cf., \cite{1367-2630-20-5-053058}) then, the relation becomes more relevant and informative. 
 Therefore, the correct interpretation of the complementarity relation is as follows: If we have large "maximum" coherence of a subsystem then the entanglement of the bipartite system will be small and vice versa.  
 
 Note that the joint state of apparatus and environment undergoes a local operation on its subsystem(apparatus), hence the entanglement $E_R(\rho_{AE})$ can never increase. Therefore, we also have the complementarity relation of the form $C_R(\rho_A)+E_R(\rho_{A^{\prime}E})\leq \log N$.
From this equation, we conclude that we should have large initial coherence of the apparatus, to keep the final entanglement between apparatus and environment small. Now, we also have a complementarity relation between final states of system and apparatus, $C_R(\rho_{A'})+E_R(\rho_{A'S'})\leq \log N$.
This relation tells us that the more the final entanglement between system and apparatus is, the less will be the final coherence of the apparatus. 

Therefore, it is clear from the above discussion and Eq.(\ref{compL1}), that to extract maximum information from a measurement process, the final entanglement between the system and apparatus should be maximum while the initial coherence of the apparatus should be as large as possible.

\subsection{Disturbance in the apparatus and information gain}
It was shown in Ref.\cite{PhysRevLett.90.050401} that the more information about a degree of freedom of a system one can gain during a measurement, the more will be the disturbance in the system. Here, we ask the opposite question: To maximize the information gain, how robust the apparatus should be? When we treat the measuring apparatus quantum mechanically, not only the apparatus disturbs the system, but also the apparatus is disturbed by the system. To answer this, we will introduce a legitimate measure of disturbance for a quantum system discussed in Ref.\cite{doi:10.1002/prop.200310045, Maccone_2007, PhysRevLett.100.210504, doi:10.1142/S1230161209000037, M_Wilde_2012}. For a quantum state $\rho_A$ evolving under a CPTP (Completely Positive and Trace Preserving) map $\Phi$ \cite{RevModPhys.86.1203}, the disturbance caused to the apparatus is given by
\begin{align*}
 D(\rho_A,\Phi)=S(\rho_A)-I_c(\rho_A,\Phi),
\end{align*}
where $I_c(\rho_A,\Phi)=\Phi(\rho_A)-S(\Phi\otimes \mathbb{I}(\ketbra{\Psi}{\Psi}_{AE})$ is the coherent information and $\ket{\Psi}_{AE}$ is the purification of $\rho_A$ such that $\rho_A={\rm Tr}_E[\ketbra{\Psi}{\Psi}_{AE}]$. The quantity $D(\rho_A,\Phi)$ satisfies the following desirable properties \cite{Maccone_2007}: (i) It should be zero iff the CPTP map is invertible on the state $\rho_A$, (ii) It should be monotonically non-decreasing under the operation of successive CPTP maps, and (iii) It should be continuous on CPTP maps and states. For completeness, we refer the reader to Ref.\cite{Maccone_2007,PhysRevA.97.062308}. This measure has been used to investigate the trade-off relation for quantum coherence and disturbance for a quantum system.

Here, to quantify the disturbance to the apparatus during the measurement process in presence of environment, we use the above quantifier. For this particular scenario, the disturbance to the apparatus caused during quantum measurement is given by
\begin{align*}
 D(\rho_A,\Phi)=S(\rho_A)-[S(\rho_{A'})-S(\rho_{A'E})].
\end{align*}
Noticing the fact that $I_m=S(\rho_{A'})-S(\rho_A)$ and $S(\rho_{A'E})\leq S(\rho_{A'})+S(\rho_{E})$, we finally have
\begin{align}
 D(\rho_A,\Phi)+I_m+C_R(\rho_{A'})\leq 2\log N
 \label{app_const}
\end{align}
where $N$ is the dimension of the apparatus and the environment. This relation is tight in the sense that the disturbance itself 
is bounded by $0\leq D(\rho_A,\Phi)\leq 2\log N$. The Eq.(\ref{app_const}) tells a very interesting feature of the apparatus itself. For a fixed amount of coherence of the apparatus final state, in order to gain more information, apparatus should be disturbed less. This is in agreement with the intuition that for maximal information gain, apparatus should be more robust during interaction with the system.

\section{Conclusion}\label{sec4}
Quantum measurement process plays a fundamental role and continues to hurl us with new insights.
We have studied the role of quantum coherence of the system and the apparatus in a quantum measurement process. We consider a measurement procedure, where the extracted information from a pure state using a mixed apparatus, is upper bounded by the initial coherence of the system. Since, we have better control over the initial coherence of the system compared to the final entanglement between apparatus and system, our result provides a realistic estimate of the extractable information. In addition, we show that, the extractable information is exactly equal to the coherent information from the final joint state of the system to the apparatus. This provides a new meaning to the information gain as the amount of the distinct quantum information that is being sent from the system to the apparatus. This finding shows us that the extractable information is rather quantum in nature than classical. We also show that the information gain by the apparatus is bounded by the entropy exchange. Further, we prove trade off relation between the information gain, disturbance, and coherence of the apparatus. To give holistic description of the measurement, we include the environment to purify the measurement apparatus and we find that to extract more information from the measurement, the apparatus should have large initial coherence and we should be able to develop maximum entanglement between the system and apparatus. The measurement procedure described here can be extended to more general measurement scenario in which the evolved apparatus states are not strictly orthogonal to each other. We hope that these findings will provide new insights to the role of coherence and coherent information in quantum measurement.
\bibliography{coholevoref}
\end{document}